\documentclass{aa}
\usepackage{graphicx}
\usepackage{epsfig}

\def\kr{\hbox{ \raisebox{-1.0mm}{$\stackrel{<}{\sim}$} }}

\begin{document}

\title{The GRB~060218/SN~2006aj event in the context of other
Gamma-Ray Burst Supernovae\thanks{Based on  observations collected at
the European Southern Observatory, Paranal, Chile  (ESO Programme
076.D-0015).}\fnmsep\thanks{Table \ref{Phottab} is only available in electronic form at 
http://www.edpsciences.org.}}

\author{
P. Ferrero\inst{1},
D. A. Kann\inst{1},
A. Zeh\inst{2}, 
S. Klose\inst{1}, 
E. Pian\inst{2}, 
E. Palazzi\inst{3},
N. Masetti\inst{3},
D. H. Hartmann\inst{4},
J. Sollerman\inst{5},
J. Deng\inst{6},
A. V. Filippenko\inst{7},
J. Greiner\inst{8},
M. A. Hughes\inst{9},
P. Mazzali\inst{2,10},
W. Li\inst{7},
E. Rol\inst{11},
R. J. Smith\inst{12},
N. R. Tanvir\inst{9,11}
}

\offprints{P. Ferrero (ferrero@tls-tautenburg.de)}


\institute{
  Th\"uringer Landessternwarte Tautenburg, D-07778 Tautenburg,
  Germany \and
  Istituto Nazionale di Astrofisica-OATs, I-34131 Trieste,
  Italy
  \and 
  INAF, Istituto di Astrofisica Spaziale e Fisica Cosmica, 
  Sez. di Bologna, I-40129 Bologna, Italy \and
  Clemson University, Department of Physics and
  Astronomy, Clemson, SC 29634-0978, USA \and
  Dark Cosmology Center, Niels Bohr Institute, Copenhagen University,
  DK-2100 Copenhagen, Denmark \and
  National Astronomical Observatories, CAS, Chaoyang District, Beijing
  100012, China \and
  Department of Astronomy, University of California,
  Berkeley, CA 94720-3411 \and
  Max-Planck-Institut f\"ur extraterrestische Physik, D-85741 Garching,
  Germany \and
  Centre for Astrophysics Research, University of Hertfordshire,
  College Lane, Hatfield, AL10 9AB, UK \and
  Max-Planck Institut f\"ur Astrophysik, D-85748
  Garching, Germany \and
  Department of Physics and Astronomy, University of Leicester,
  Leicester, LE1 7RH, UK \and
  Astrophysics Research Institute, Liverpool John Moores University,
  Twelve Quays House, Birkenhead, CH41 1LD
}

\date{Received: May 2006 / Accepted: }

\authorrunning {Ferrero et al.}

\titlerunning{SN~2006aj and GRB supernovae}


\abstract{The supernova SN~2006aj associated with GRB 060218 is the 
second-closest GRB-SN observed to date ($z$=0.033). We present Very Large
Telescope, Liverpool Telescope, and Katzman Automatic Imaging Telescope
multi-color photometry of SN~2006aj. This supernova is found to
be subluminous and rapidly evolving. Its early light curve includes an 
additional wavelength-dependent component, which can be interpreted
as shock break-out. We compare the photometric evolution of multi-band
light curves with the corresponding properties of the present sample of 
more than 10 GRB-SNe with precisely known redshifts. Using host-galaxy 
extinction measurements, we derive extinction-corrected GRB-SN luminosities 
and place SN~2006aj in the context of this GRB-selected supernova 
sample as well as in the context of local stripped-envelope supernovae. 

}

\maketitle

\section{Introduction}

There is now substantial evidence for the association of long-duration
GRBs with core-collapse supernovae (SNe of Type Ic to be more specific;
see Filippenko \cite{Filippenko1997} for a review of supernova spectral 
classification). 
The first hint for such a relation came with the contemporaneous
discovery of GRB980425 and of a local SN (1998bw) in its error circle
(Galama et al. \cite{Galama1998}).  Subsequently, Bloom et
al. (\cite{Bloom1999}) reported a possible underlying SN component in
the afterglow of GRB 980326. The first spectroscopic identification of
a SN~Ic superposed on a GRB afterglow (GRB~030329/SN~2003dh; Hjorth et
al. \cite{Hjorth2003}, Matheson et al. \cite{Matheson2003}, Stanek et
al. \cite{Stanek2003}) solidified this association considerably.

The photometric database on GRB-SNe was summarized by Zeh, Klose \&
Hartmann (\cite{PaperI, Texas}, henceforth Z04, Z05) who, by studying 
a sample of 21 (Z04) and 29 (Z05) bursts with well established redshifts
in a systematic way, found or confirmed a weak SN component for 9
(Z04) and 13 (Z05)  sources in those sets, respectively.
In particular, all long bursts
with redshifts $z\kr0.7$ were found to have a late-time bump in their
optical afterglows, suggesting that in fact all long bursts are
physically associated with SN explosions. Additional photometric and
partly spectroscopic results on recent GRB-SNe with well-known
redshift were presented by Malesani et al. (\cite{Malesani2004};
GRB 031203/SN~2003lw), Soderberg et al. (\cite{Soderberg2006}) and Stanek et
al. (\cite{Stanek2005}) on SNe associated with GRB 040924 and GRB
041006, as well as Della Valle et al. (\cite{DellaValle2006};
GRB 050525A/SN~2005nc). A recent review of the status of the supernova - GRB connection is
presented by Woosley \& Bloom (\cite{WoosleyBloom2006}).

GRB 980425/SN~1998bw is still the closest GRB-SN to date, and poses 
the question whether it is a typical representative of this class, a
``standard source'' in some sense. Its comparatively low isotropic energy
release in the gamma-ray band (less than 10$^{48}$ erg, Galama et
al. \cite{Galama1998}) makes this burst stand out from the
cosmological GRB ensemble with known redshifts (cf. Ghirlanda et
al. \cite{Ghirlanda2004}). It is therefore of great interest to find
additional events in the nearby universe. Consequently, the discovery
of GRB 060218 (Cusumano et al. \cite{Cusumano2006}) (more precisely an
X-Ray Flash, XRF) and its identification with an energetic SN by
Masetti et al. (\cite{Masetti2006}), now designated SN~2006aj
(Soderberg, Berger, \& Schmidt 2006), attracted much attention (Modjaz
et al. \cite{Modjaz2006}, Sollerman et al. \cite{Sollerman2006},
Mirabal et al. \cite{Mirabal2006}, Cobb et al. \cite{Cobb2006}).  Its
small distance ($z=0.033$, Mirabal \& Halpern
\cite{MirabalHalpernGCN}) results in an isotropic equivalent energy of
the burst of $5\,\times 10^{49}$ erg in the 15.5-154.8 keV band, which
again is underenergetic in comparison to the cosmological GRB sample
(Modjaz et al. \cite{Modjaz2006}), whose mean redshift is larger than
$z$=1 (Jakobsson et al. \cite{Jakobsson2006}, Bagoly et
al. \cite{Bagoly2006}).

Here, we report results from a photometric ESO Very Large
Telescope (VLT) campaign on GRB~060218/SN~2006aj covering a time span
of nearly 4 weeks, with additional data gathered with the robotic
Liverpool Telescope (LT) at La Palma and the robotic Katzman Automatic
Imaging Telescope (KAIT; Filippenko et al. \cite{Filippenko2001},
Li et al. \cite{Li2003a}) at Lick Observatory. Results
from  the spectroscopic VLT campaign are presented in Pian et
al. (\cite{Pian2006})  and modeled in Mazzali et
al. (\cite{Mazzali2006}). While previous publications on SN~2006aj
discussed the X-ray and optical light curves in detail, we place the
photometric evolution of SN~2006aj in the context of all presently
known GRB-SNe with precise redshifts. In particular, we update the
fits reported in Z04 and Z05, add several more recent
GRB-SNe, and correct the photometry for host-galaxy extinction as
derived in Kann, Klose \& Zeh (\cite{PaperIII}, henceforth K06). 

\section{Observations and data reduction}

\subsection{VLT data}

We observed SN~2006aj both spectroscopically and photometrically with
the ESO Very Large Telescope FORS1 and FORS2 instruments. Observations
were performed until 26 days after the burst (Table \ref{Phottab}). After 
this point in time, the SN location was no longer observable
due to high airmass. For the photometry the exposure time was between
30 and 60 seconds.  The images, in the Bessell $B$, $V$, $R$, and $I$ filters,
were bias-subtracted and flat-fielded
with standard reduction procedures using IRAF\footnote{
http://iraf.noao.edu} and final photometry was performed with standard
Point Spread Function (PSF) fitting using the DAOPHOT II image data
analysis package (Stetson \cite{Stetson1987}) within
MIDAS\footnote{http://www.eso.org/projects/esomidas}. Photometric
calibration of the images was performed using the standard star 685 of
the field SA98 (Landolt \cite{Landolt1992}), imaged during the same night of our first
observational run, which was used to create 16 secondary standards in
the field around the SN that were subsequently used for all individual
images to draw a linear regression among instrumental and reference
magnitudes and calculate the value of the optical transient. The
accuracy of the photometry was confirmed by considering the zero
point values of several stars in the calibration field in all bands for
different nights. In addition, we investigated a possible influence of 
color terms. We found that they could affect the photometry with an
additional error of up to 0.02 magnitudes, which was then added in
quadrature to the individual photometric measurement errors.

\subsection{Liverpool Telescope data and KAIT data}

Further photometric data were obtained with the 2~m robotic Liverpool Telescope
on La Palma on several occasions over a period of about 2.5 weeks post-burst. 
Observations were made in Bessell $B$, $V$ and SDSS $r^{\prime}$ filters and 
zero-point reduced by comparison to two field stars (star S1 and star
4 of Modjaz et al. \cite{Modjaz2006}). The magnitudes were transformed
into Johnson $B$, $V$ and Cousins $R$-band via equations previously derived for
the camera (Steele \cite{Steele2004}). The quoted
errors reflect the scatter in individual measurements for subexposures at
each epoch, and include an estimate of the calibration uncertainty.

Additional data were also obtained with the 0.8 m robotic KAIT at Lick
Observatory on four consecutive nights between 3 to 6 days post-burst.
Unfiltered observations were obtained in all four nights, and Bessell
$B$, $V$, $R$, and $I$  filtered observations were made in the last
three nights. The images were bias-subtracted and flat-fielded with
standard reduction procedures using IRAF, and the final photometry was
performed with PSF-fitting technique in IRAF/DAOPHOT. The secondary
standards calibrated with VLT data were used for the final photometric
calibration. The unfiltered magnitudes were converted to the $R$ band
following the method described in Li et al. (2003b).

\begin{figure}
\includegraphics[width=8.5cm,angle=0,clip=true]{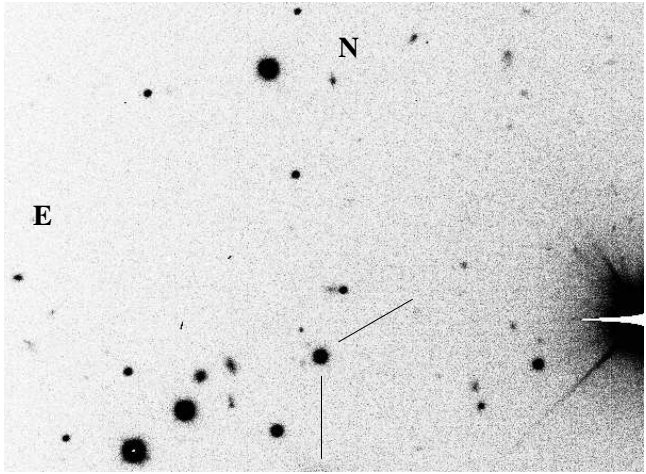}
\caption{The location of SN~2006aj is shown here in a 
VLT/FORS2 $R$-band image taken on 2006 March 1, at 00:52 UT (30 sec exposure time).
The field of view is approximately 2$\times$3 arcmin.}
\label{field}
\end{figure}

\subsection{The magnitudes of the host galaxy}

The host galaxy of GRB~060218/SN~2006aj was imaged pre-burst by the
Sloan Digital Sky Survey (Cool et al. \cite{Cool2006},
Adelman-McCarthy et al. \cite{SDSSV4}). It was pointed out (Modjaz et
al. \cite{Modjaz2006}, Hicken et al. \cite{Hicken2006}) that there are
offsets between the SDSS calibration and later efforts, rendering the
transformed host-galaxy magnitudes of Cool et al. (\cite{Cool2006})
too bright.  The values used for the host galaxy in this paper are
$B=20.57\pm0.07$, $V=20.18\pm0.04$, $R=20.03\pm0.03$ and
$I=19.58\pm0.06$, not corrected for Galactic extinction.  These
magnitudes have been evaluated from those reported by Cool et
al. (\cite{Cool2006}), by applying the corrections recommended by
Modjaz et al. (\cite{Modjaz2006}) and Hicken et
al. (\cite{Hicken2006}). Our $BVRI$ magnitudes have been obtained from
Cool et al.'s $u^*b^*v^*r^*i^*$ magnitudes by adopting the conversion
of Lupton (2005)\footnote{see also
http://www.sdss.org/dr4/algorithms/sdssUBVRI
Transform.html\#Rodgers2005} and taking into account that the offsets
determined by Modjaz et al. (2006) in the $BVR$ bands are 0.40, 0.27,
0.20 mag, respectively\footnote{see
http://www.cfa.harvard.edu/oir/Research/supernova/
sn2006aj/compstars.html}.  In order to determine the offset in the $I$
band, we have considered that  the offset determined by Hicken et
al. (\cite{Hicken2006}) and Modjaz et al. (\cite{Modjaz2006}) in  the
$i^*$-filter is 0.15 mag, and we have used the SDSS
transformations\footnote{see
http://www.sdss.org/dr4/algorithms/jeg\_photometric
\_eq\_dr1.html\#usno2SDSS} to derive from this the offset in the
$i$-band. We note that our host-galaxy magnitudes are consistent to those
of Sollerman et al. (\cite{Sollerman2006}) within the errors.
As the host galaxy and SN 2006aj were not separated at the resolution 
of the images, the host magnitudes have to be subtracted.

\subsection{The extinction correction}

As the host galaxy and SN 2006aj were not separated at the resolution 
of the images, the host magnitudes have to be subtracted.
For the extinction correction,
we used the values derived by Guenther et al. (\cite{Guenther2006})
from an analysis of the Na~I~D2 absorption lines in our Galaxy
($A_V=0.39$ mag) and the host galaxy ($A_V=0.13$ mag) in a VLT UVES
spectrum, obtained during this campaign close to SN maximum. We note
that this host-galaxy extinction is typical of the lines of sight
toward GRBs (K06). Using these extinction and host magnitude values,
we derive the magnitudes of the pure SN light component. 

\begin{table*}[h!]
\caption{
The photometry of SN~2006aj from the Very Large Telescope (VLT), the
Liverpool Telescope (LT) and the Katzman Automatic Imaging Telescope
(KAIT), including the acquisition images for the VLT spectroscopy
(Pian et al. \cite{Pian2006}). The date is the UT exposure
mid-time. Column $t-t_0$ refers to the time in days after the burst
trigger at $t_0$ = 3:34:30 UT of Feb. 18, 2006 (Cusumano et
al. \cite{Cusumano2006}). The third column gives measured magnitudes
which are corrected neither for extinction nor flux from the host galaxy. 
The fourth column gives the pure supernova magnitudes after correcting for 
Galactic extinction, host-galaxy flux and host-galaxy extinction. Exposure 
times are in seconds.}
\begin{center}
\begin{tabular}{rrccccl}
\hline
Date & $t-t_0$ & Mag. (measured) & Mag. (corrected) & Exposure & Filter & Telescope \\ \hline
February	20.8924	&	2.7434	& $18.63\pm0.06$ & $18.14\pm0.06$ &	8$\times$150	&	B	&	LT	\\
February	21.0307	&	2.8817	& $18.57\pm0.03$ & $18.07\pm0.03$ &    2$\times$30	&	B	&	VLT	\\
February	22.1219	&	3.9729	& $18.41\pm0.08$ & $17.87\pm0.08$ &	450	&	B	&	KAIT	\\
February	23.0550	&	4.9060	& $18.24\pm0.04$ & $17.68\pm0.04$ &    2$\times$30	&	B	&	VLT	\\
February	23.1209	&	4.9719	& $18.29\pm0.08$ & $17.74\pm0.08$ &	450	&	B	&	KAIT	\\
February	24.1218	&	5.9728	& $18.13\pm0.08$ & $17.56\pm0.08$ &	450	&	B	&	KAIT	\\
February	25.0467	&	6.8977	& $18.11\pm0.04$ & $17.54\pm0.04$ &    2$\times$30	&	B	&	VLT	\\
February	26.0398	&	7.8908	& $18.05\pm0.03$ & $17.47\pm0.03$ &    2$\times$30	&	B	&	VLT	\\
February	27.0483	&	8.8993	& $18.01\pm0.03$ & $17.43\pm0.03$ &	 60	&	B	&	VLT	\\
February	28.0037	&	9.8547	& $18.07\pm0.04$ & $17.49\pm0.04$ &    2$\times$30	&	B	&	VLT	\\
March	1.0337	&	10.8847	& $18.14\pm0.03$ & $17.57\pm0.03$ &	2$\times$30	&	B	&	VLT	\\
March	2.0297	&	11.8807	& $18.22\pm0.03$ & $17.66\pm0.03$ &	2$\times$30	&	B	&	VLT	\\
March	3.0330	&	12.8840	& $18.35\pm0.03$ & $17.81\pm0.03$ &     2$\times$30	&	B	&	VLT	\\
March	4.0303	&	13.8813	& $18.47\pm0.03$ & $17.95\pm0.03$ &	2$\times$30	&	B	&	VLT	\\
March	5.0119	&	14.8629	& $18.61\pm0.03$ & $18.11\pm0.03$ &	2$\times$30	&	B	&	VLT	\\
March	6.0364	&	15.8874	& $18.66\pm0.03$ & $18.17\pm0.03$ &	2$\times$30	&	B	&	VLT	\\
March	7.8899	&	17.7409	& $18.97\pm0.07$ & $18.56\pm0.07$ &	5$\times$100	&	B	&	LT	\\
March	8.0332	&	17.8842	& $19.03\pm0.03$ & $18.64\pm0.04$ &	  60	&	B	&	VLT	\\
March	8.9983	&	18.8493	& $19.10\pm0.03$ & $18.73\pm0.04$ &	  60	&	B	&	VLT	\\
March	10.0009	&	19.8519	& $19.26\pm0.04$ & $18.95\pm0.05$ &	  60	&	B	&	VLT	\\
March	11.9977	&	21.8487	& $19.40\pm0.04$ & $19.16\pm0.05$ &    60+30	&	B	&	VLT	\\
March	13.0145	&	22.8655	& $19.57\pm0.05$ & $19.43\pm0.07$ &	  60	&	B	&	VLT	\\
March	13.9976	&	23.8486	& $19.57\pm0.05$ & $19.43\pm0.07$ &	  60	&	B	&	VLT	\\ \hline
February	20.8787	&	2.7279	& $18.19\pm0.06$ & $17.86\pm0.06$ &	5$\times$150	&	V	&	LT	\\
February	21.0323	&	2.8833	& $18.17\pm0.03$ & $17.83\pm0.03$ &	2$\times$30	&	V	&	VLT	\\
February	22.1277	&	3.9787	& $17.92\pm0.08$ & $17.54\pm0.08$ &	300	&	V	&	KAIT	\\
February	23.0573	&	4.9083	& $17.80\pm0.03$ & $17.41\pm0.03$ &	2$\times$30	&	V	&	VLT	\\
February	23.1266	&	4.9776	& $17.77\pm0.08$ & $17.37\pm0.08$ &	300	&	V	&	KAIT	\\
February	24.1275	&	5.9785	& $17.69\pm0.08$ & $17.28\pm0.08$ &	300	&	V	&	KAIT	\\
February	24.8713	&	6.7223	& $17.61\pm0.06$ & $17.19\pm0.06$ &	5$\times$100	&	V	&	LT	\\
February	25.0483	&	6.8993	& $17.58\pm0.03$ & $17.16\pm0.03$ &	2$\times$30	&	V	&	VLT	\\
February	26.0413	&	7.8923	& $17.51\pm0.03$ & $17.08\pm0.03$ &	2$\times$30	&	V	&	VLT	\\
February	26.8720	&	8.7230	& $17.48\pm0.06$ & $17.05\pm0.06$ &	5$\times$100	&	V	&	LT	\\
February	27.0156	&	8.8666	& $17.45\pm0.03$ & $17.02\pm0.03$ &	60	&	V	&	VLT	\\
February	27.0497	&	8.9007	& $17.46\pm0.03$ & $17.03\pm0.03$ &	60	&	V	&	VLT	\\
February	28.0053	&	9.8563	& $17.45\pm0.03$ & $17.02\pm0.03$ &	2$\times$30	&	V	&	VLT	\\
March	1.0353	&	10.8863	& $17.45\pm0.03$ & $17.02\pm0.03$ &	2$\times$30	&	V	&	VLT	\\
March	2.0313	&	11.8823	& $17.47\pm0.03$ & $17.04\pm0.03$ &	2$\times$30	&	V	&	VLT	\\
March	2.8673	&	12.7183	& $17.50\pm0.06$ & $17.07\pm0.06$ &	3$\times$100	&	V	&	LT	\\
March	3.0345	&	12.8855	& $17.51\pm0.03$ & $17.08\pm0.03$ &	2$\times$30	&	V	&	VLT	\\
March	4.0331	&	13.8841	& $17.56\pm0.03$ & $17.14\pm0.03$ &	2$\times$30	&	V	&	VLT	\\
March	5.0134	&	14.8644	& $17.60\pm0.03$ & $17.18\pm0.03$ &	2$\times$30	&	V	&	VLT	\\
March	6.0379	&	15.8889	& $17.68\pm0.03$ & $17.27\pm0.03$ &	2$\times$30	&	V	&	VLT	\\
March	6.8912	&	16.7422	& $17.67\pm0.08$ & $17.26\pm0.08$ &	2$\times$100	&	V	&	LT	\\
March	8.0018	&	17.8528	& $17.85\pm0.03$ & $17.46\pm0.03$ &	60	&	V	&	VLT	\\
March	8.0347	&	17.8857	& $17.86\pm0.03$ & $17.47\pm0.03$ &	60	&	V	&	VLT	\\
March	8.9998	&	18.8508	& $17.92\pm0.03$ & $17.54\pm0.03$ &	60	&	V	&	VLT	\\
March	9.0067	&	18.8577	& $17.92\pm0.03$ & $17.54\pm0.03$ &	60	&	V	&	VLT	\\
March	10.0024	&	19.8534	& $18.01\pm0.03$ & $17.65\pm0.03$ &	60	&	V	&	VLT	\\
March	10.0080	&	19.8590	& $17.98\pm0.03$ & $17.61\pm0.03$ &	60	&	V	&	VLT	\\
March	11.9987	&	21.8497	& $18.14\pm0.03$ & $17.80\pm0.03$ &	60+30	&	V	&	VLT	\\
March	13.0157	&	22.8667	& $18.22\pm0.04$ & $17.89\pm0.04$ &	60	&	V	&	VLT	\\
March	13.9991	&	23.8501	& $18.30\pm0.03$ & $17.99\pm0.03$ &	60	&	V	&	VLT	\\
March	14.9992	&	24.8502	& $18.41\pm0.03$ & $18.12\pm0.03$ &	2$\times$30	&	V	&	VLT	\\
March	16.0023	&	25.8533	& $18.42\pm0.05$ & $18.14\pm0.05$ &	2$\times$30	&	V	&	VLT	\\ \hline
\end{tabular}
\end{center}
\label{Phottab}
\end{table*}


\begin{table*}[h!]
\addtocounter{table}{-1}
\caption{Table \ref{Phottab} continued}
\begin{center}
\begin{tabular}{rrccccl}
\hline
February	20.8578	&	2.7088	& $18.10\pm0.06$ & $17.88\pm0.06$ &	6$\times$150	&	R	&	LT	\\
February	20.9362	&	2.7872	& $18.10\pm0.06$ & $17.88\pm0.06$ &	6$\times$150	&	R	&	LT	\\
February	21.0339	&	2.8849	& $18.02\pm0.03$ & $17.78\pm0.03$ &	2$\times$30	&	R	&	VLT	\\
February	21.0374	&	2.8884	& $18.02\pm0.03$ & $17.78\pm0.03$ &	30	&	R	&	VLT	\\
February	21.1294	&	2.9804	& $17.98\pm0.08$ & $17.73\pm0.08$ &	120	&	R	&	KAIT	\\
February	21.1312	&	2.9822	& $17.96\pm0.08$ & $17.71\pm0.08$ &	120	&	R	&	KAIT	\\
February	21.1330	&	2.9840	& $17.97\pm0.08$ & $17.73\pm0.08$ &	120	&	R	&	KAIT	\\
February	21.1349	&	2.9859	& $17.98\pm0.08$ & $17.73\pm0.08$ &	120	&	R	&	KAIT	\\
February	21.1367	&	2.9877	& $17.96\pm0.09$ & $17.71\pm0.09$ &	120	&	R	&	KAIT	\\
February	22.1317	&	3.9827	& $17.76\pm0.08$ & $17.48\pm0.08$ &	300	&	R	&	KAIT	\\
February	22.1398	&	3.9908	& $17.79\pm0.09$ & $17.51\pm0.09$ &	120	&	R	&	KAIT	\\
February	23.0224	&	4.8734	& $17.66\pm0.03$ & $17.37\pm0.03$ &	60	&	R	&	VLT	\\
February	23.0590	&	4.9100	& $17.66\pm0.03$ & $17.37\pm0.03$ &	2$\times$30	&	R	&	VLT	\\
February	23.1306	&	4.9816	& $17.60\pm0.08$ & $17.30\pm0.08$ &	300	&	R	&	KAIT	\\
February	23.1387	&	4.9897	& $17.60\pm0.10$ & $17.29\pm0.10$ &	120	&	R	&	KAIT	\\
February	24.1315	&	5.9825	& $17.50\pm0.09$ & $17.19\pm0.09$ &	300	&	R	&	KAIT	\\
February	24.1395	&	5.9905	& $17.50\pm0.10$ & $17.19\pm0.10$ &	120	&	R	&	KAIT	\\
February	24.8645	&	6.7155	& $17.44\pm0.07$ & $17.12\pm0.07$ &	5$\times$100	&	R	&	LT	\\
February	25.0499	&	6.9009	& $17.40\pm0.04$ & $17.08\pm0.04$ &	2$\times$30	&	R	&	VLT	\\
February	26.0429	&	7.8939	& $17.31\pm0.04$ & $16.98\pm0.04$ &	2$\times$30	&	R	&	VLT	\\
February	26.8677	&	8.7187	& $17.29\pm0.06$ & $16.96\pm0.06$ &	3$\times$100	&	R	&	LT	\\
February	27.0512	&	8.9022	& $17.28\pm0.03$ & $16.95\pm0.03$ &	60	&	R	&	VLT	\\
February	28.0068	&	9.8578	& $17.25\pm0.03$ & $16.91\pm0.03$ &	2$\times$30	&	R	&	VLT	\\
March	1.0370	&	10.8880	& $17.23\pm0.03$ & $16.89\pm0.03$ &	2$\times$30	&	R	&	VLT	\\
March	2.0328	&	11.8838	& $17.23\pm0.03$ & $16.89\pm0.03$ &	2$\times$30	&	R	&	VLT	\\
March	2.8604	&	12.7114	& $17.26\pm0.06$ & $16.93\pm0.06$ &	5$\times$100	&	R	&	LT	\\
March	3.0360	&	12.8870	& $17.24\pm0.03$ & $16.90\pm0.03$ &	2$\times$30	&	R	&	VLT	\\
March	3.8601	&	13.7111	& $17.25\pm0.07$ & $16.91\pm0.07$ &	2$\times$100	&	R	&	LT	\\
March	4.0346	&	13.8856	& $17.26\pm0.03$ & $16.93\pm0.03$ &	2$\times$30	&	R	&	VLT	\\
March	5.0148	&	14.8658	& $17.24\pm0.04$ & $16.90\pm0.04$ &	2$\times$30	&	R	&	VLT	\\
March	6.0082	&	15.8592	& $17.33\pm0.03$ & $17.00\pm0.03$ &	60	&	R	&	VLT	\\
March	6.0394	&	15.8904	& $17.33\pm0.03$ & $17.00\pm0.03$ &	2$\times$30	&	R	&	VLT	\\
March	6.8834	&	16.7344	& $17.36\pm0.07$ & $17.03\pm0.07$ &	4$\times$100	&	R	&	LT	\\
March	7.8841	&	17.7341	& $17.41\pm0.06$ & $17.09\pm0.06$ &	5$\times$100	&	R	&	LT	\\
March	8.0361	&	17.8871	& $17.46\pm0.03$ & $17.14\pm0.03$ &	60	&	R	&	VLT	\\
March	9.0012	&	18.8522	& $17.56\pm0.03$ & $17.26\pm0.03$ &	60	&	R	&	VLT	\\
March	10.0038	&	19.8548	& $17.60\pm0.03$ & $17.30\pm0.03$ &	60	&	R	&	VLT	\\
March	11.9997	&	21.8507	& $17.62\pm0.04$ & $17.32\pm0.04$ &	60+30	&	R	&	VLT	\\
March	13.0169	&	22.8679	& $17.68\pm0.03$ & $17.39\pm0.03$ &	60	&	R	&	VLT	\\
March	14.0013	&	23.8523	& $17.89\pm0.03$ & $17.63\pm0.03$ &	60	&	R	&	VLT	\\ \hline
February	21.0358	&	2.8868	& $17.86\pm0.03$ & $17.80\pm0.03$ &	2$\times$30	&	I	&	VLT	\\
February	22.1357	&	3.9867	& $17.69\pm0.08$ & $17.60\pm0.08$ &	300	&	I	&	KAIT	\\
February	23.0604	&	4.9114	& $17.54\pm0.03$ & $17.41\pm0.03$ &	2$\times$30	&	I	&	VLT	\\
February	23.1346	&	4.9856	& $17.49\pm0.09$ & $17.36\pm0.09$ &	300	&	I	&	KAIT	\\
February	24.1356	&	5.9866	& $17.31\pm0.09$ & $17.15\pm0.09$ &	300	&	I	&	KAIT	\\
February	25.0514	&	6.9024	& $17.25\pm0.03$ & $17.08\pm0.03$ &	2$\times$30	&	I	&	VLT	\\
February	26.0444	&	7.8954	& $17.16\pm0.03$ & $16.98\pm0.03$ &	2$\times$30	&	I	&	VLT	\\
February	27.0526	&	8.9036	& $17.10\pm0.03$ & $16.91\pm0.03$ &	60	&	I	&	VLT	\\
February	28.0083	&	9.8593	& $17.05\pm0.04$ & $16.85\pm0.04$ &	2$\times$30	&	I	&	VLT	\\
March	1.0386	&	10.8896	& $17.04\pm0.03$ & $16.84\pm0.03$ &	2$\times$30	&	I	&	VLT	\\
March	2.0328	&	11.8854	& $17.03\pm0.03$ & $16.83\pm0.03$ &	2$\times$30	&	I	&	VLT	\\
March	3.0376	&	12.8886	& $17.03\pm0.03$ & $16.83\pm0.03$ &	2$\times$30	&	I	&	VLT	\\
March	4.0362	&	13.8872	& $17.07\pm0.04$ & $16.87\pm0.04$ &	2$\times$30	&	I	&	VLT	\\
March	5.0165	&	14.8675	& $17.01\pm0.06$ & $16.81\pm0.06$ &	2$\times$30	&	I	&	VLT	\\
March	6.0410	&	15.8920	& $17.08\pm0.03$ & $16.89\pm0.03$ &	2$\times$30	&	I	&	VLT	\\
March	9.0026	&	18.8536	& $17.21\pm0.03$ & $17.03\pm0.03$ &	60	&	I	&	VLT	\\
March	10.0052	&	19.8562	& $17.26\pm0.03$ & $17.09\pm0.03$ &	60	&	I	&	VLT	\\
March	11.9993	&	21.8503	& $17.35\pm0.05$ & $17.19\pm0.05$ &	60+30	&	I	&	VLT	\\
March	13.0181	&	22.8691	& $17.37\pm0.05$ & $17.21\pm0.05$ &	60	&	I	&	VLT	\\
March	14.0027	&	23.8537	& $17.48\pm0.03$ & $17.34\pm0.03$ &	60	&	I	&	VLT	\\ \hline
\end{tabular}
\end{center}
\end{table*}

\section{Results}

We followed previous works (Z04, Z05) and modeled  the light curve
of SN2006aj
using SN~1998bw as a template (Galama et al. \cite{Galama1998}) that
was shifted\footnote{Throughout the paper we adopt $H_0=71$ km
s$^{-1}$ Mpc$^{-1}$, $\Omega_{\rm M}=0.27$, $\Omega_\Lambda=0.73$
(Spergel et al. \cite{Spergel2003}), which for $z$=0.033 yields a
distance modulus of 35.78 mag.} to the corresponding redshift and
scaled in luminosity (in the SN  rest frame) by a stretch factor $k$ and in
time evolution by a factor $s$, while zero host extinction was assumed
for SN~1998bw (Patat et al. \cite{Patat2001}). In doing so, we find an
additional component visible in the early data that makes the light
curve systematically brighter than that of SN~1998bw. This component
has also been noted by Mirabal et al. (\cite{Mirabal2006}).
Furthermore, strong spectral evolution is noticeable, with the flux
excess being stronger toward shorter wavelengths. The most reasonable
explanation for this additional blue component is the light due to
shock break-out through a dense progenitor wind (Campana et
al. \cite{Campana2006}). However, we caution that at least part of this additional component may be the result of an intrinsically different (as compared to SN 1998bw) early SN light curve. SN 2002ap (cf. Appendix) is also overluminous at early times in comparison to SN 1998bw, and has been compared, both photometrically and spectroscopically, with SN 2006aj (e.g. Pian et al. \cite{Pian2006}). But the early multi-color evolution of SN 2002ap is achromatic, showing no evidence for strong spectral evolution as seen in SN 2006aj.

\begin{figure}
\includegraphics[width=8.5cm,angle=0,clip=true]{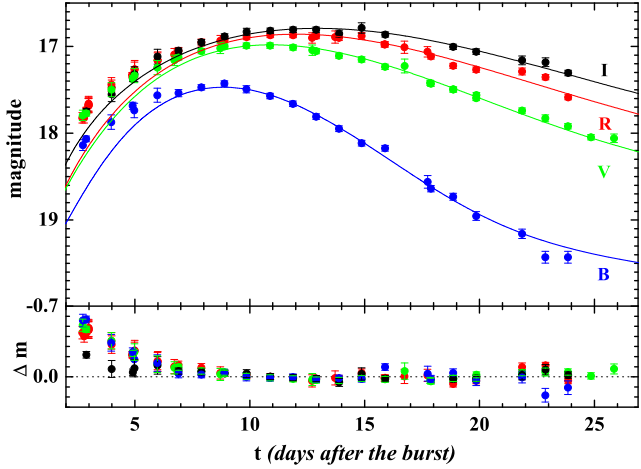}
\caption{
The light curves of SN~2006aj based on our $BVRI$ data after correcting 
for extinction and host flux contribution. The data were fitted 
using the light curves of SN~1998bw as a template, not considering 
data from SN~2006aj that was taken
before 8.8 days post burst. The residuals $\Delta$m represent observed 
values minus the fit.}
\label{LC1}
\end{figure}

We thus used two different methods to fit the SN light curve.
In the first fit, we excluded all data earlier than 8.8 days, but fit 
only the SN 1998bw model light curve to the data, without any additional 
contributions. The result is shown in Fig. \ref{LC1}. The early excess 
blue flux is clearly visible in the residuals. The data after 8.8 days 
are matched very well by the SN 1998bw model. Note that for typical GRB-SNe
we also fit only the peak and decaying part of the SN, as the early 
flux is typically dominated by the GRB afterglow.

\begin{figure}
\includegraphics[width=8.5cm,angle=0,clip=true]{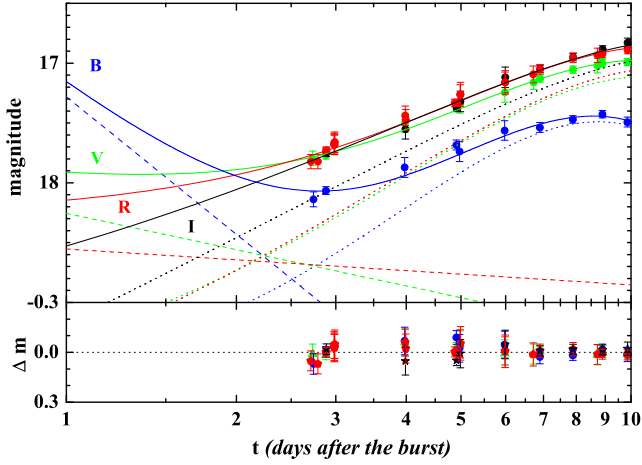}
\caption{
The light curves of SN~2006aj fitted by including an additional early
component which decays according to a power-law.  Solid lines mark the
fit, the dashed lines are the power-law component and the dotted lines
are the SN component. The time scale is logarithmic to show in detail
the earlier data.}
\label{LC2}
\end{figure}

In the second fit, we assumed that an additional component decays according 
to a power-law, with the decay index $\alpha$ ($F_{\nu}\propto
t^{-\alpha}$) being a free parameter of the fit, and independent for
each band (Fig. \ref{LC2}). The derived values of the luminosity ratio
$k$ and the stretch factor $s$ as well as the decay index $\alpha$ are
given in Table \ref{ksresults}. It is apparent that while the results
derived from the two methods are similar, they do in fact differ by as
much as 10\% (in $k$). The additional component improves the quality of
the fit in all bands. The steeper decay at shorter wavelengths is also in
agreement with the \emph{Swift} UVOT photometry of the blue shock
break-out (Campana et al. \cite{Campana2006}). The strong wavelength
dependence of the decay (Table \ref{ksresults}) argues against this
component being the optical afterglow of the GRB. On the other hand, the 
very shallow decay in $VRI$ indicates that the time evolution of the SED of 
this additional component is probably more complex than a simple decay according 
to a pure power-law, as assumed in our analysis.

\begin{table}[htb]
\caption{Fitting results for the light curves of SN~2006aj attained 
by either excluding early data (top) or including an additional 
power-law component decaying with index $\alpha$ (see
text for details).}
\begin{center}
\begin{tabular}{llccr}
\hline
	&	$\chi^2_\nu$	&	$k$	&	$s$	&	$\alpha$	\\ \hline
$B$	&	1.83	&	$0.734\pm0.014$	&	$0.620\pm0.006$	&	$\cdots$	\\
$V$	&	1.02	&	$0.724\pm0.007$	&	$0.682\pm0.005$	&	$\cdots$	\\
$R$	&	1.93	&	$0.735\pm0.007$	&	$0.689\pm0.006$	&	$\cdots$	\\
$I$	&	0.71	&	$0.761\pm0.008$	&	$0.686\pm0.008$	&	$\cdots$	\\ \hline
$B$	&	1.79	&	$0.721\pm0.022$	&	$0.613\pm0.005$	&	$1.53\pm0.43$	\\
$V$	&	0.35	&	$0.640\pm0.014$	&	$0.659\pm0.008$	&	$0.40\pm0.09$	\\
$R$	&	0.94	&	$0.621\pm0.022$	&	$0.656\pm0.012$	&	$0.12\pm0.10$	\\
$I$	&	0.48	&	$0.667\pm0.039$	&	$0.663\pm0.022$	&	$-0.10\pm0.34$	\\ \hline
\hline
\end{tabular}
\end{center}
\label{ksresults}
\end{table}

\begin{table}[htb]
\caption{
Luminosity ratio $k$ and stretch factor $s$ for GRB-SNe \emph{with known redshift} 
normalized to SN~1998bw before and after
correcting for host extinction. In cases where no extinction could be
derived, only the non-corrected values are given. In all cases the SN
fits were derived in the observer $R$-band frame. This list is
complete up to the end of 2005, with the exception of GRB 040924
(Soderberg et al. \cite{Soderberg2006}). 
These results supersede those presented in Z05. Differences
compared to Z05 are basically due to the inclusion of new
observational data or a revision of the used value for the
Galactic extinction along the line of sight.
Furthermore, note
that we use different cosmological parameters than those in Z04 and Z05.}
\begin{center}
\begin{tabular}{lccc}
\hline
GRB/XRF/SN	&	$k$					&	$s$					&	$k$ (corrected)								\\ \hline
970228	        &	$	0.40\pm0.29	$	&	$	1.45\pm0.95	$	&	$\cdots$								\\
990712	        &	$	0.35\pm0.09	$	&	$	0.83\pm0.13	$	&	$\cdots$								\\
991208	        &	$	0.93\pm0.25	$	&	$	1.12\pm0.20	$	&	$	2.62^{+1.10}_{-0.65}	$	\\
000911	        &	$	0.63\pm0.29	$	&	$	1.40\pm0.32	$	&	$	0.85^{+0.44}_{-0.26}	$	\\
010921	        &	$	0.68\pm0.43	$	&	$	0.69\pm0.25	$	&	$	1.85^{+2.82}_{-0.79}	$	\\
011121/2001ke	  &	$	0.59\pm0.02	$	&	$	0.80\pm0.02	$	&	$	0.88^{+0.08}_{-0.07}	$	\\
020405	        &	$	0.74\pm0.05	$	&	$	0.97\pm0.07	$	&	$	0.90^{+0.15}_{-0.11}	$	\\
020903	        &	$	0.62\pm0.09	$	&	$	0.92\pm0.08	$	&	$\cdots$								\\
021211/2002lt	  &	$	0.40\pm0.19	$	&	$	0.98\pm0.26	$	&	$\cdots$								\\
030329/2003dh	  &	$	1.06\pm0.11	$	&	$	0.85\pm0.10	$	&	$	1.50^{+0.19}_{-0.16}	$	\\
031203/2003lw	  &	$	0.67\pm0.05	$	&	$	1.09\pm0.07	$	&	$	1.28^{+0.18}_{-0.16}	$	\\
041006	        &	$	0.90\pm0.05	$	&	$	1.38\pm0.06	$	&	$	1.03^{+0.22}_{-0.09}	$	\\
050525A/2005nc	&	$	0.49\pm0.04	$	&	$	0.77\pm0.04	$	&	$	0.66^{+0.10}_{-0.08}	$	\\ \hline
\end{tabular}
\end{center}
\label{GRBSN}
\end{table}

The small value of the stretch factor $s$ means that SN~2006aj evolves
much faster than SN~1998bw (see also Pian et al. \cite{Pian2006}, Modjaz et
al. \cite{Modjaz2006}, Sollerman et al. \cite{Sollerman2006}, Mirabal
et al. \cite{Mirabal2006}, Cobb et al. \cite{Cobb2006}, Soderberg et al. 
\cite{Soderberg2006c}). Furthermore,
SN~2006aj has a slightly different color than SN~1998bw, being
brighter in the $B$ and $I$ band than in the $V$ and $R$ band.  The
$k$ values we obtain are also in accordance with the values found by
the authors cited above. In all bands, SN~2006aj is about 30\% less
luminous than SN~1998bw. From the light curve data, independent of any
fit, we find the following peak  absolute magnitudes of the SN:
$M_B=-18.29\pm0.05$, $M_V=-18.76\pm0.05$, $M_R=-18.89\pm0.05$,
$M_I=-18.95\pm0.10$.

In order to place SN~2006aj in the context of the presently known GRB-SNe, 
we started with the results presented in Z04 and Z05, updated them in 
cases where new data were available, and corrected the $k$ value using 
host-galaxy extinctions derived in K06, including uncertainties. 
We added two further GRB-SNe to the sample. The result for XRF 020903 is
based on data from Bersier et al. (\cite{Bersier2006}). We found
that the SN of this event is also superimposed on an underlying
power-law afterglow component. Unfortunately, the afterglow data are
too sparse to derive any conclusions on the extinction in the host
galaxy. For GRB 031203, we used the extinction derived by Mazzali et
al. (\cite{Mazzali031203}). In the case of GRB 050525A, we employed the data
presented by Della Valle et al. (\cite{DellaValle2006}) and applied the
extinction found by Blustin et al. (\cite{Blustin2006}). In those
cases where no extinction could be found in K06, we only present the
observed values and note that these are lower limits on the luminosity
of the SNe (Table \ref{GRBSN}). On the other hand, the line of sight
extinction is often low and the true values are thus not expected to
be much higher.

\section{Discussion}

\subsection{SN~2006aj in the context of known GRB-SNe}

SN~2006aj is a rare nearby GRB-SN, with a distance of only about 140
Mpc. Its luminosity is about 70\% of that of of SN~1998bw, confirming
that GRB-selected supernovae may not qualify for the label ``standard 
candle''. Given the UVES data of SN~2006aj, taken close to SN maximum 
light (Guenther et al. \cite{Guenther2006}, Wiersema et al. \cite{Wiersema2006}), 
it is unlikely that the deficiency in luminosity is the result of dust 
extinction in the host galaxy of SN~2006aj. 

\begin{figure}
\includegraphics[width=8.5cm,angle=0,clip=true]{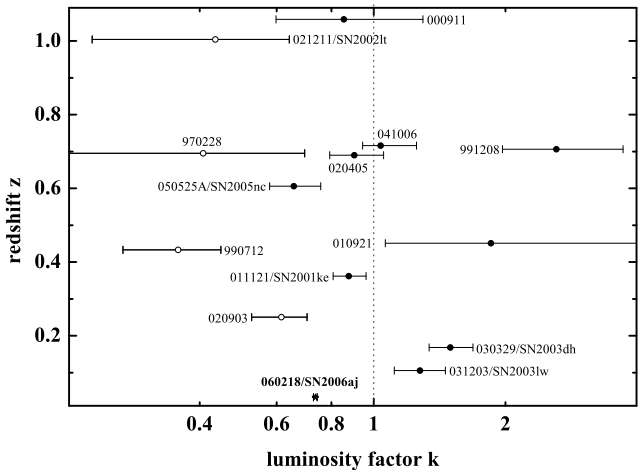}
\caption{
The distribution of the luminosity factor $k$ in the $R$ band with
redshift $z$. There is no apparent correlation with redshift, even
though we caution that all the $k$ values refer to the $R$ band in the
observer frame, corresponding to different wavelengths in the host
frame. Filled symbols mark the $k$ values that have been corrected
for host extinction, open symbols represent supernovae for which this
correction could not be applied. A star marks the value derived from the
$R$-band light curve of SN~2006aj by removing data before 8.8 days
from the fit. SN~1998bw is at $k=1$, by definition.}
\label{kz}
\end{figure}

\begin{figure*}
\includegraphics[width=18.1cm,angle=0,clip=true]{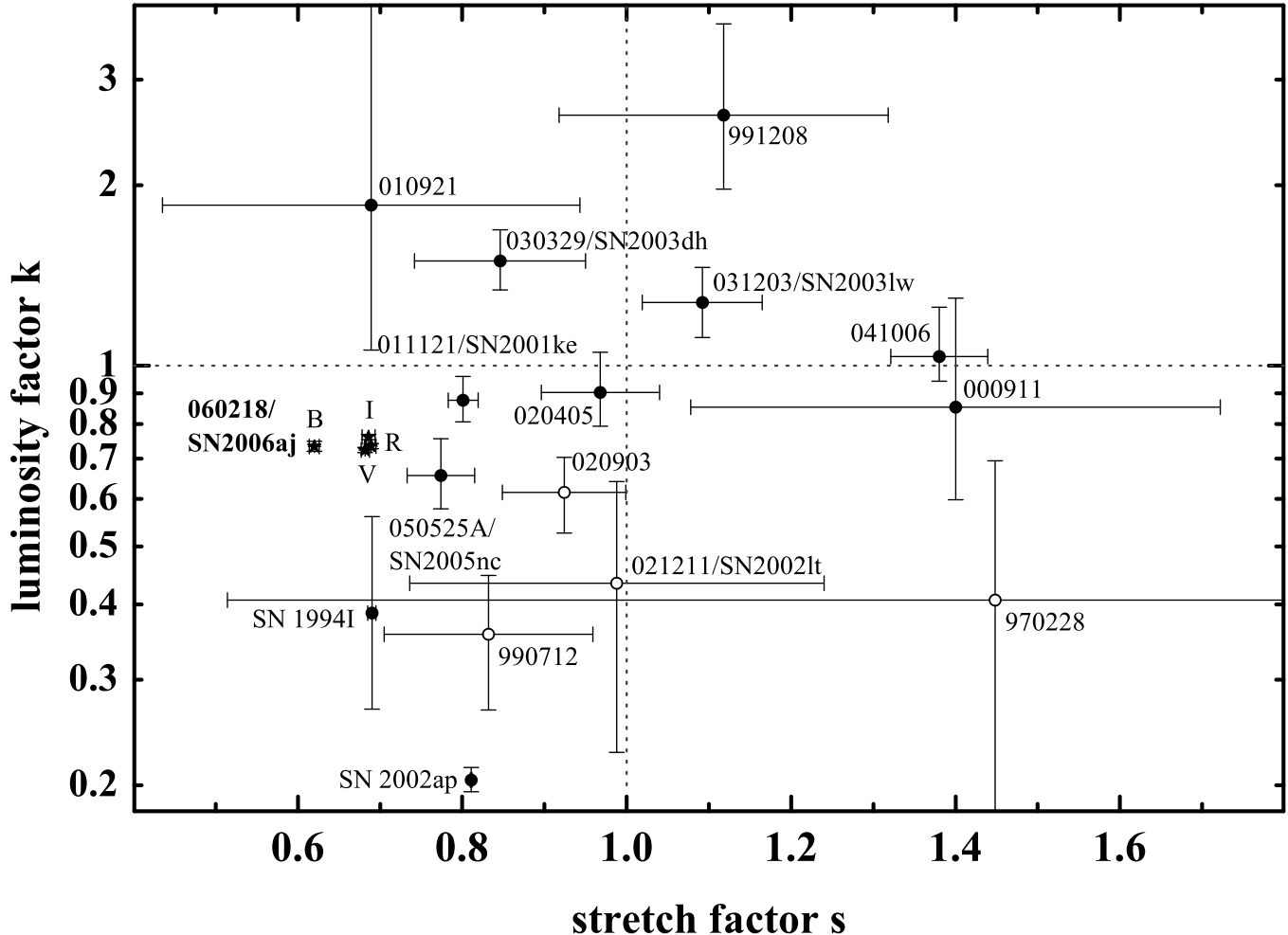}
\caption{
Luminosity factor $k$ versus stretch factor $s$ in the $R$ band for
all GRB-SNe in our sample (Table \ref{GRBSN}). The symbols are
identical to those used in Fig. \ref{kz}. SN~1998bw defines $k=1$,
$s=1$. Note that in the case of SN~2006aj the values for $BVRI$ are plotted.
Furthermore, we plot the Type Ic supernova SN 1994I and the Type Ic broad-lined SN Ic
SN 2002ap, neither of which are associated with GRBs. Both are fainter than any
GRB-SN in our sample for which the host extinction has been corrected.}
\label{ks}
\end{figure*}

By the end of 2005, there were 14 optical afterglows\footnote{Compared
to Z04/Z05, GRB 980703 has been removed as we now have reason to doubt 
the identification of a SN bump.} with known redshifts that
showed evidence for extra light in the $R$ band at late times (GRBs
970228, 990712, 991208, 000911, 010921, 011121, 020405, 021211,
030329, 031203, 040924\footnote{This event is not included in our
study as sufficient photometry has not been published yet.},
041006, 050525A, XRF 020903). For eight afterglows with a known SN
bump (GRBs 991208, 000911, 010921, 011121, 020405, 021211, 030329,
041006) K06 were able to derive a host extinction value. We used these
to determine the extinction-corrected luminosities  of these SNe in
the observer frame (Fig. \ref{kz}). It turns out that only two
(030329, 031203) or perhaps three (including 991208) of the 14 SNe
were actually more luminous than SN~1998bw with high significance
(but see, e.g., Deng et al. \cite{Deng2005}, for a spectroscopic
modeling of SN~2003dh that results in $k<1$). Remarkably, most
extinction corrected SNe cluster around $0.6<k<1.5$. We find that the
four SNe not corrected for extinction (GRBs 970228, 990712, 020903,
021211) are typically fainter, implying that a correction for host
galaxy extinction will probably also shift them into this range. While 
one might worry about the fact that for each individual GRB-SN, due to 
the different redshifts, the $k$ factor refers to a different wavelength 
region, SN~2006aj does not contradict our assumption that $k$ is not 
strongly dependent on wavelength (Fig. \ref{ks}).  

We conclude that the present data, even though some have large
uncertainties, indicate that the width of the GRB-SN luminosity
function is at least 2 mag, comparable to what  is known for the other
types of SNe (cf. Richardson et al. \cite{Richardson2002}, Richardson,
Branch \& Baron \cite{RBB2006}, henceforth RBB06). In particular,
there is no evidence that the luminosity function evolves with
redshift: The width of the luminosity function for $z<0.2$ is
comparable to the width at $z\approx 0.7$. It is also much narrower
than the distribution of intrinsic afterglow luminosities (K06).

\subsection{SN~2006aj in the context of local stripped-envelope supernovae}

\begin{figure}
\includegraphics[width=8.8cm,angle=0,clip=true]{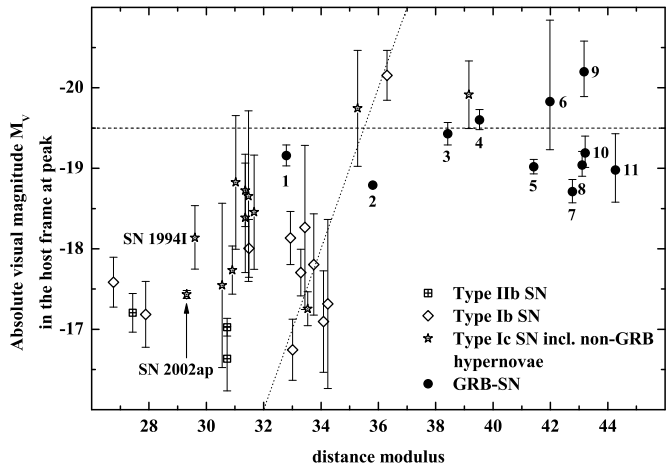}
\caption{
The absolute $V$-band magnitude $M_V$ of stripped-envelope supernovae
versus their distance modulus. This is
an expanded version of Figure 1 in RBB06, and the values for SN of Type
IIb, Ib and Ic have been taken from that work but transformed
according to the world model used here.
The individual GRB-SNe have the same numbers as
in Fig.~\ref{mMAV}. We have also labeled two local Type Ic SNe we have
included in our study (Fig.~\ref{ks}; Appendix). 
GRB-SNe without known host extinction are not included. The
dashed line at $M_V=-19.5$ denotes the typical absolute magnitude of
Type Ia SNe (the ``ridgeline'').  The slanted dotted line denotes a
constant visual magnitude $m_V=16$ in the case of no extinction.  }
\label{mMMV}
\end{figure}

It is also interesting to compare the light curve properties of
SN 2006aj and other GRB-SNe with well-studied local stripped-envelope 
(i.e. types Ib, Ic, IIb) supernovae, as is shown in Fig.~\ref{mMMV}. 
Distance moduli and absolute
magnitudes for these SNe were taken from RBB06 (who use  $H_0=60$ km
s$^{-1}$ Mpc$^{-1}$) and have been transformed to the world model used
here ($H_0=71$ km s$^{-1}$ Mpc$^{-1}$) by adding $-5 \log 71/60 =-
0.365$ mag to the former and +0.365 mag to the latter values.
Similarly, for the world model used here SN 1998bw is 0.19 mag less
luminous than given in Galama et al. (1998), who use $H_0=65$ km
s$^{-1}$ Mpc$^{-1}$, i.e. $k$=1 corresponds to $M_V = - 19.16$. The
absolute visual magnitudes of the GRB-SNe in our sample were then
calculated according to  $M_V^{GRB-SN} - M_V^{98bw} = - 2.5 \log k$,
assuming that $k$ is independent of wavelength. Figure~\ref{mMMV}
shows the result obtained in this way. Most notable is that the
ensemble of GRB-SNe is at the bright end of the luminosities of local
stripped-envelope supernovae. In other words, \it the present data
indicate that Type Ic supernovae with associated (detected) GRBs are
on average more luminous in the optical bands than those without detected
GRBs. In particular, SN 2006aj is no exception from this rule. \rm We
caution, however, that our assumption of wavelength-independence of
$k$ could be an oversimplification. The larger the redshift, the more 
uncertain is the absolute $V$-band magnitude of a GRB-SN derived in this 
way. This uncertainty is not included in the error bars plotted in 
Fig.~\ref{mMMV}.

Finally, we note that an observational bias might affect
the interpretation of Fig.~\ref{mMMV}: the larger the redshift, the more difficult it is to observe
less luminous stripped-envelope supernovae.

\begin{figure}
\includegraphics[width=8.8cm,angle=0,clip=true]{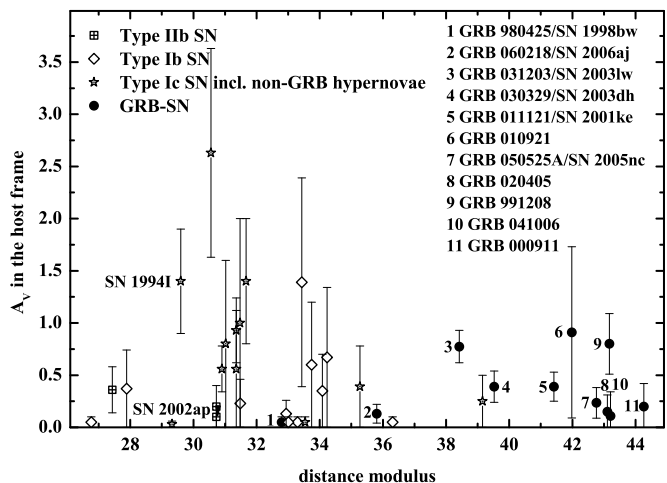}
\caption{
The visual host-galaxy extinction $A_V$ (in the host frame) of
stripped-envelope supernovae versus their distance modulus. The values 
for SNe of Type IIb, Ib and Ic have been taken from RBB06, while 
the data for the GRB-SNe are from K06. We labeled individual GRB-SNe 
and the two Type Ic SNe we have also studied. GRB-SNe without known host 
extinction were not included.
}
\label{mMAV}
\end{figure}

RBB06 give host extinction values for all 27 events in their sample. 
Therefore, we investigated whether host extinction of GRB-SNe is 
different from those of local stripped-envelope supernovae. No 
substantial differences are apparent (Fig.~\ref{mMAV}). 
While it is interesting that SN 2006aj is less affected by host 
extinction compared to the local sample of Type Ic SNe, the current
sample is too small to draw reliable conclusions from this finding.

\subsection{The stretch factor}

Finally, we consider the statistics of the stretch factor $s$, which
is in some sense more reliable since it is not affected by the
extinction issue. First of all, among all known GRB-SNe, SN~2006aj
has the smallest $s$ value, i.e. it is the fastest GRB-SN ever seen
(Pian et al. \cite{Pian2006}, Mazzali et al. \cite{Mazzali2006},
Modjaz et al. \cite{Modjaz2006}). Interestingly,  about half of the
GRB-SNe have $s<1$, i.e., they are evolving faster than SN~1998bw. In
contrast, the evolution of the SN associated with GRB 041006 was very
slow (Fig. \ref{ks}) with high significance (Stanek et
al. \cite{Stanek2005}). Basically only two SNe (000911, 970228) occupy
the ($k<1$, $s>1$) region, i.e., they are slow and subluminous,
contrary to the general trend, but both have large error bars. Stanek
et al. (\cite{Stanek2005}) noted that for GRB-SNe a relation may exist
between light curve shape and luminosity, similar to the one established 
for Type Ia SNe (see Figure 3 in Stanek et al. \cite{Stanek2005}). Fig.~\ref{ks} neither 
supports nor contradicts the existence of such a potential relation. 
While a trend of rising $k$ with rising $s$ seems to be visible, a fit 
to the data does not support this trend with reasonable statistical significance.

\subsection{Shortcomings of the procedure}

The procedure applied here has shortcomings that are basically related 
to cosmological effects on one hand and to data quality on the other.

As long as one is only concerned with broad-band photometry, which is 
the case in this study, the observed light curves of
GRB-SNe usually refer to different wavelength bands in their host
frames. However for SN 1998bw it has been shown that the light curve shape, the
time of the peak flux, and the peak flux itself are a function of
wavelength (as is the case for other SN types, too). Unfortunately,
the available photometric data base is in most cases restricted to
$R$-band observations (in the observer frame). In order to be able to
compare light curves of GRB-SNe that occurred at various
redshifts with light curves of SN 1998bw, the simplest and in some
sense only useful approach is then to assume that the luminosity
ratio $k$ and the stretch factor $s$, which are both normalized to SN
1998bw, are independent of wavelength. In other words, we assume that
the SED of all GRB-SNe (in their host frame) is the same at all times.

In addition, the data base used for our light curve fits is usually
weighted to data obtained past the peak time of a GRB-SN under
consideration. The reason is that before the peak time usually
the afterglow dominates the light of the optical transient, while
after the peak time usually the SN light dominates. This problem is
difficult to overcome. Therefore, for basically all cosmologically
remote SNe we have no information about the details of the SN light
curves at early times up to several days after the corresponding
burst. In particular the stretch factor $s$ is then mainly affected by
the late-time behaviour of the SN light curves.

It is therefore by no means clear if a relation between $s$ and $k$ 
indeed exists. It is possible that such a relation is hidden by the 
relatively large error bars for individual ($k,s$) values of the GRB-SNe 
on the one hand (Fig.~\ref{ks}) and by the basic assumptions that went 
into the procedure we applied here on the other hand. More spectral data 
from GRB-SNe might finally solve this issue. However, progress made in 
this regard during the last years was only modest, at best. Most Swift 
detected GRBs are at such a high redshift that no SN spectroscopy can be 
performed with current telescopes within a reasonable amount of observing 
time. Therefore, the photometric approach utilized here derives some of
its justification from these spectroscopic limitations.

\section{Conclusions}

SN~2006aj is the fastest evolving and one of the least luminous GRB-SN
discovered so far, being only about 70\% as luminous as
SN~1998bw. This places it at the faint end  of the GRB-SN luminosity
distribution which so far, after extinction correction, covers the
range from about 0.6 to 2 times the luminosity of SN~1998bw. Placing
SN 2006aj in the context of the luminosities of local Type Ic
supernovae without detected GRB (RBB06) reveals, however, that SN
2006aj is still at the bright end of their luminosity distribution.
SN 2006aj thus follows the general ``rule'' that GRB-SNe tend to be
more luminous than (local) Type Ic SNe without detected GRB.

GRBs 980425 and 060218 suggest that the prompt emission properties
are not correlated with the optical properties of the associated SNe. 
GRB 980425 had a duration of 31 sec (Pian et al. \cite{Pian1999}), 
while GRB 060218 lasted for more than 2000 sec (Campana et al. 
\cite{Campana2006}), and the corresponding isotropic equivalent energy 
release in the gamma-ray band was different by a factor of 10. The 
corresponding SN luminosities, however, differ by only 30\%. Assuming 
that differences in instrumental sensitivities (BeppoSAX versus Swift) 
do not play a role here, this supports the notion that the properties of
the GRB and the associated SN are to a large extent independent of
each other.

\begin{acknowledgements}

We thank the staff at ESO/Paranal, in particular A. Bik, C. Dumas,
O. Hainaut,  E. Jehin, C. Ledoux, O. Marco, C. Melo, D. Naef,
L. Schmidtobreick  for performing the observations.  F.P., D.A.K. and
S.K. acknowledge support by DFG grant Kl 766/13-2 and by the German
Academic Exchange Service (DAAD) under grant No. D/05/54048.
A.V.F. is supported by NASA/Swift grants  NNG05GF35G and NNG06GI86G. KAIT 
and its ongoing research were made possible by generous donations from Sun
Microsystems, Inc., the Hewlett-Packard Company, AutoScope Corporation, Lick
Observatory, the National Science Foundation, the University of California,
the Sylvia \& Jim Katzman Foundation, and the TABASGO Foundation.
This work has benefitted from collaboration within the EU FP6 Research 
Training Network HPRN-CT-2002-00294 ``Gamma-Ray Bursts: an Enigma and a 
Tool.'' We thank the anonymous referee for a prompt and constructive report.

\end{acknowledgements}

\appendix
\section{SN~1994I and SN~2002ap}

Following the suggestion of the referee, we consider here two
local Type Ic supernovae that were not physically related to a
detected GRB but are often used for comparison with GRB-SNe, and place
them in the $k,s$ context used in the present paper. While the fit
appears to be worse, it should be noted that when fitting a GRB-SN in
all cases the fit is practically always weighted to data points
lying after the light curve maximum when the SN light dominates the
afterglow light.

\emph{SN 1994I} in the Whirlpool Galaxy M~51 (Puckett et
al. \cite{Puckett1994}) counts as the ``prototypical'' Type Ic
supernova (Filippenko et al. \cite{Filippenko1995}), 
although it is faster than any other (RBB06). For the
analysis of the light curve we used $R$-band data from Yokoo et
al. (\cite{Yokoo1994}), Schmidt \& Kirschner (\cite{CFA}), Lee et al.
(\cite{Lee1995}) and Richmond et al. (\cite{Richmond1996}) (with
late data taken only from Schmidt \& Kirschner and Richmond et al.).
SN 1994I occurred near the bulge of M~51 and was highly
reddened: Richmond et al. estimate $E_{(B-V)}=0.45\pm0.16$ with a
Milky Way  extinction curve. The explosion date of the supernova is
not well known. From light curve modeling, Iwamoto et
al. (\cite{Iwamoto1994}) derive an explosion date of JD 2449433. These
authors equate this with a $B$-band maximum 12 days after the
explosion. A comparison with the data from Richmond et
al. (\cite{Richmond1996}) shows, however, that JD 2449437 corresponds
to 12 days before the $B$ maximum. Using the latter, the early light
curve becomes strongly subluminous in comparison to SN 1998bw, has to
be excluded from the fit and $s=0.69\pm0.01$ is found from the
location of the maximum alone. 
On the other hand, if we take March 31.0 UT (JD 2449442.5)
as the explosion date, we find that up to an age of 25 days the early
light curve is well fitted by the SN~1998bw light curve. Assuming a
distance modulus of 29.60 mag (Richmond et al. \cite{Richmond1996}),
corresponding to a distance  $d$=8.34 Mpc for the world model used
here, we find, after correcting for the host-galaxy extinction
(Richmond et al. \cite{Richmond1996}), $k=0.39^{+0.17}_{-0.12}$.  Thus
SN 1994I is fainter than any GRB-SN when we account for host extinction.

\begin{figure}
\includegraphics[width=8.8cm,angle=0,clip=true]{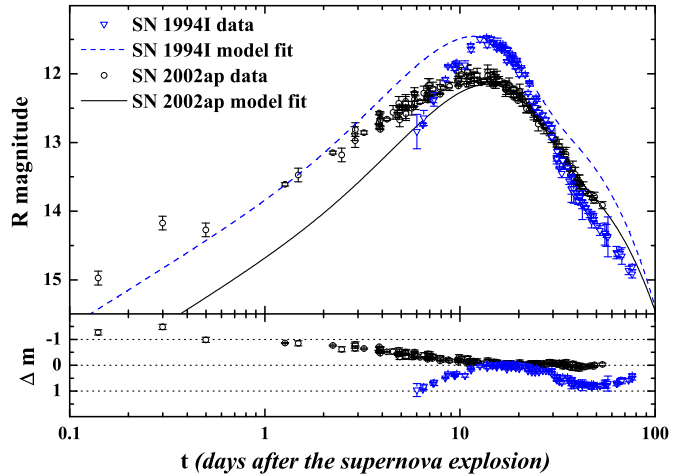}
\caption{
The two local Type Ic supernovae SN~1994I and SN~2002ap fitted
with the model light curve of SN~1998bw at the respective
redshifts. In both cases, the light curve of SN 1998bw does not fit the
light curves of the other SNe completely. The best agreement is obtained
for SN 2002ap, where data after ten days up to 60 days are fitted well by
the SN 1998bw light curve. For SN 1994I data before and
after the maximum had to be excluded from the fit.}
\label{SNe}
\end{figure}

\emph{SN 2002ap}, which exploded in the outer spiral arms of M~74 (Nakano
et al. \cite{Nakano2002}), is the closest and best followed-up
broad-lined SN Ic to date. A search through data of all IPN satellites did not
reveal any GRB that could be associated with the SN (Hurley et
al. \cite{Hurley2002}); still, the explosion date was well
constrained, as the supernova was discovered very rapidly (Mazzali et
al. \cite{Mazzali2002}). For the analysis of the light curve we
collected data from the following sources: Gal-Yam et al.
(\cite{Galyam2002}), Borisov et al. (\cite{Borisov2002}), Cook et
al. (\cite{Cook2002}), Nishihara et al. (\cite{Nishihara2002}), Yoshii
et al. (\cite{Yoshii2003}), Pandey et al. (\cite{Pandey2003}),
Doroshenko et al. (\cite{Doroshenko2003}), Foley et al.
(\cite{Foley2003}), and Vink\'o et al. (\cite{Vinko2004}). In addition we
made use of the data collected at
SNWEB\footnote{http://www.astrosurf.com/snweb2/2002/02ap/02apMeas.htm}.
Based on these data SN 2002ap is overluminous at early times in
comparison to SN 1998bw. The host-galaxy extinction is very small,
$E_{(B-V)}=0.020\pm0.0008$  (Takada-Hidai et al. 2002).  Correcting
for this and the Galactic extinction, and assuming a distance modulus
of 29.32 mag (Foley et al. \cite{Foley2003}), which corresponds to a
distance $d$=7.3 Mpc for the cosmological model used here, we find
$k=0.219\pm0.0002$ and $s=0.747\pm0.0006$. Thus, the
stretch factor $s$ for SN 2002ap is very close to what we find for SN
2006aj, but the SN is much fainter.


\end{document}